\documentclass[aps,prb,twocolumn,superscriptaddress,raggedbottom,showpacs,floatfix]{revtex4}
\usepackage{amsmath,amsfonts,amssymb,amsthm}
\usepackage{color}
\usepackage{graphicx,latexsym}
\usepackage{eurosym}
\usepackage{dcolumn}
\usepackage{bm}

\begin{document}

\title
{\bf  Modulated currents in open nanostructures}

\author{Valeriu Moldoveanu}
\affiliation{National Institute of Materials Physics, P.O. Box MG-7,
Bucharest-Magurele, Romania}
\author{Andrei Manolescu}
\affiliation{School of Science and Engineering, Reykjavik University, Kringlan 1, 103 Reykjavik, Iceland }
\author{Vidar Gudmundsson}
\affiliation{Science Institute, University of Iceland, Dunhaga 3, IS-107 Reykjavik, Iceland}
\begin{abstract}

We investigate theoretically the transport properties of a mesoscopic system driven by  
a sequence of rectangular pulses applied at the contact to the input (left) lead.   
The characteristics of the current which would be measured in the output (right) lead 
are discussed in relation with the spectral properties of the sample. 
The time-dependent currents are calculated via a generalized non-Markovian master equation scheme. 
We study the transient response of a quantum dot and of a narrow quantum wire. 
We show that the output response depends not only on the lead-sample coupling and on the  
length of the pulse but also on the states that propagate the input signal. 
We find that by increasing the bias window the new states available for transport 
induce additional structure in the relaxation current due to different dynamical 
tunneling processes. The delay of the output signal with respect to the input current in 
the case of the narrow quantum wire is associated to the transient time through the wire. 

\end{abstract}

\pacs{73.23.Hk, 85.35.Ds, 85.35.Be, 73.21.La}

\maketitle

\section{Introduction}

Time-dependent transport measurements at nanoscale provide important
insight into the intrinsic properties of semiconductor structures like
relaxation and dephasing times \cite{Fujisawa} and play a crucial role
in single-shot spin read-out schemes. \cite{Hanson} Consequently
transient  response of quantum structures to pulsed signals gains interest
from experimental point of view. Recently Naser {\it et al.} \cite{Naser}
measured the time-dependent current through a quantum point
contact when a pulse generator is coupled to the
input lead. The output signal was measured for different amplitudes
and rise times of the pulse. The decay of the output current was roughly
exponential and it was suggested that such a device could be used as
a microwave circuit element.  In another recent experiment Lai {\it
et al.} \cite{Lai} performed transient current measurements for
a Ge quantum dot when trapezoidal voltage pulses were applied to one
electrode. The transient current mimics the pulse shape and
the extracted time-dependent occupation number in the system follows an
accumulation/depletion cycle.

From the theoretical point of view the transient currents have been
calculated by various methods. Stefanucci {\it et al.} \cite{Stefanucci1,
Stefanucci2} combined the Keldysh formalism and the density functional
theory to study the response of a nanostructure to a time-dependent bias
between the leads. In the partitioning approach the transient currents
induced  through a few-level quantum dot by a modulation of the contacts
between the leads and the sample was calculated in Ref.\onlinecite{Moldo1}
and a scattering theory of the time-dependent magnetotransport
in a long quantum wire was proposed in Ref.\onlinecite{Gudmundsson}.
%On the other hand Pedersen and Wacker \cite{Pedersen} solved the 
%Liouville equation up to the second order in the tunneling Hamiltonian.

These studies were essentially focused on understanding the transient
regime and the onset of the steady-state. Therefore the time-dependent
driving used in the calculations did not describe pulsed signals,
but rather an initial switching stage followed by a constant,
time-independent value.  In particular sudden coupling or smooth
coupling lead to qualitatively different output signals.  Another problem
considered in the time-dependent transport calculations is the quantum
pumping. \cite{Switkes,Arrachea,Splettstoesser,Stefanucci3} In this
case one obtains averaged currents through an unbiased system which is
perturbed by {\it two} time-dependent potentials oscillating out-of-phase.

In this work we discuss transport calculations for a mesoscopic sample
driven out of equilibrium by both a constant bias applied between the
leads and a fast oscillating signal applied at the contact between
the lead and the sample.  If the signal has a rectangular shape the
setup is in some sense similar to the pump-and-probe configuration
used in the transient spectroscopy experiments of Fujisawa {\it et al.}
\cite{Fujisawa}  The time-dependent signal was applied on the sample and
the main aim was to extract the spin relaxation time by pushing {\it one}
excited state into the transport window during the pulse.

Here we discuss the transient response of the system from a different
angle: If a sequence of rectangular pulses that modulate the coupling
to the left lead is viewed as an input signal then one can study
the propagation of this signal trough the system and the corresponding
current in the right (i.e.\,output) lead.  Our problem is therefore closer
to the experiments by Naser {\it et al.} \cite{Naser} and Lai {\it et
al.} \cite{Lai} Besides the very ambitious goal of assembling quantum
dot structures in complex mesoscopic circuits operating like quantum
gates we believe there is another important motivation for such a study.
The transient response of the sample to the modulation of the contact is
a consequence of the internal electron dynamics which depends crucially
on the electronic states participating in transport. The point is that
by changing the bias applied on the system one selects different states
in the transport window and then the output current may carry important
informations about the electron dynamics in the sample.  Even if the level
structure of the sample may be known from other type of measurements,
the tunneling rates associated to each quantum state or the propagation
properties of the 'orbitals' are not easily understood.

We aim at describing the following transport experiment through a
mesoscopic structure: the contact between the sample and the left lead 
opens and closes periodically by applying rectangular pulses on the
metallic gates that define the contact region. At the same time the contact 
between the sample and the right lead gradually opens.  Then each time the left
contact closes the electrons in the sample can only escape into the right
lead. These relaxation processes of various states in the sample depend
mainly on the corresponding tunneling rates which in turn are given by
the coupling of the states to the output contact.  In real samples the
states that participate in transport have different tunneling coefficients
and therefore it is not obvious that the relaxation current follows a
simple exponential decay.  We show that the transient response of the
sample is in general more complex.  Sometimes the output current may
look exponential, but often may also carry a fine structure reflecting
the presence of several quantum states propagating through the sample
and various relaxation processes corresponding to transitions between
these states and the right lead.

A recent attempt to describe rectangular pulses in quantum dots was
made by Oh {\it et al.} \cite{Oh} Their sample model was a two-level
quantum dot and for transport calculations they used rate equations
with time-dependent tunneling coefficients.  Instead, our calculations
are done using the generalized  non-Markovian master equation
(GME) approach presented in one of our recent works. \cite{Moldo2}
We take into account the geometry of the system and its spectral
properties which determine the tunneling coefficients and therefore
the currents driven by the external bias.  The application of the
GME method to quantum transport received attention in the last
years in the context of transient currents, measurement theory or
full-counting statistics or coherent control of transport (see e.g.
Refs.\,\onlinecite{Harbola,LiX,Vaz,Rammer,Braggio,Urban,Welack1,Welack2}).

The content of the paper is organized as follows: the formalism is
briefly explained in Section II, the numerical simulations are reported
in Section III, and the main conclusions are included in Section IV.

\section{The Generalized master Equation method}

In this Section we introduce the Hamiltonian of the system, the
underlying notations, and we summarize the main equations derived
in our previous work Ref.\onlinecite{Moldo2}.  We consider
a mesoscopic system which is coupled to two leads at $t=0$
and described by the following second-quantized Hamiltonian
($h.c.$ denotes Hermitian conjugation): 
\begin{eqnarray}\nonumber
H(t)&=&\sum_{l=L,R}\int dq\varepsilon^l(q)c_{ql}^{\dagger}c_{ql}+\sum_n
E_nd_n^{\dagger}d_n\\\label{HHat} &+&\sum_{l=L,R}\sum_n\int
dq\chi^l(t) (T^l_{qn}c_{ql}^{\dagger}d_n+h.c), 
\end{eqnarray} 
where $c_{ql}^{\dagger}$, $c_{ql}$, and $d_n^{\dagger}$, $d_n$ are
creation and destruction operators for electrons with momentum $q$
and energy $\varepsilon^l(q)$ in the lead $l$, and with energy $E_n$
in the sample, respectively. The labels $L$ and $R$ denote the left and
right lead. The third term is the tunneling Hamiltonian
and contains the time-dependent switching functions $\chi^l(t)$  and
the coupling matrix elements $T_{qn}^{L,R}$ associated to each pair of
states $\{\psi_q^{L/R},\phi_n\}$ from the leads and the sample. 
The pulses applied at the contact region between the left lead and 
the sample are simulated by the function $\chi_L$ which by construction has a 
rectangular shape. 
 Although our formalism could also be implemented for a continuous model (see
Ref.\,\onlinecite{Gudmundsson2}) here we use a lattice model for which
the matrix elements are given by (see Ref.\,(\onlinecite{Moldo2})):
\begin{equation}\label{Tqn} T^{l}_{qn}=V_l{\psi}^{l*}_q(0_l)\phi_n(i_l),
\end{equation} 
where $V_l$ is the coupling strength between the lead $l$ and the sample,
$0_l$ is the site of the lead $l$ which couples to the contact site
$i_l$ in the sample. The eigenfunctions of the sample $\phi_n$ and the
corresponding energies $E_n$ are numerically computed while those in
the leads, $\psi^l_q$ and $\varepsilon^l(q)$ are known analytically:
\begin{equation}\label{psiq}
\psi^l_q(m)=\frac{\sin(q(m+1))}{\sqrt{2t_L\sin q}},\quad
\varepsilon^l(q)=2t_L\cos q ,  
\end{equation} 
$t_L$ being the hopping constant on leads.  We emphasize that even in this
simple lattice model the coupling matrix elements introduced in Eq.(\ref{Tqn})
depend both on the energy and of the localization of the
sample states. A similar model for the transfer Hamiltonian was proposed
by Maddox {\it et al.} \cite{Maddox}

The statistical operator of the open quantum system is denoted by $W$
and solves the Liouville equation: $i\dot W(t)=[H(t),W(t)]$, with
the initial condition $W(t=0)=\rho_L \rho_R \rho_S$.  This means
that before the coupling the sample is described by the statistical
operator $\rho_S$ defined just below, and the leads are characterized by
equilibrium distributions $\rho_{L,R}$ with different chemical potentials
$\mu_L>\mu_R$.

The many-body states of the system are described by the sequence of occupation
numbers of the single-particle states $\{\phi_n\}$ for the isolated
system.  We shall denote the many-body states by Greek letters, i.e.
$|{\bf \nu } \rangle = |i_1^{\nu},i_2^{\nu},..,i_n^{\nu}...\rangle $
and by $i_n^{\nu}$ the occupation number of the $n$-th single particle
state. If the initial state of the disconnected sample is $|\nu_0\rangle$ then 
$\rho_S=|\nu_0\rangle\langle {\bf \nu}_0|$. 
For example, if two electrons are situated on the lowest
levels at $t \leq 0$ we have $|\nu_0\rangle=|110....\rangle$.

Following the main lines of the superoperator method \cite{Haake,Timm}
we take the partial trace of $W$ over the Fock space of the leads and end
up with a master equation for the reduced density operator 
$\rho(t)={\rm Tr}_L {\rm Tr}_R W(t)$ with the initial condition $\rho(t=0)=\rho_S$
up to second order in the tunneling Hamiltonian:
\begin{eqnarray}\nonumber
{\dot\rho}(t)&=&-\frac{i}{\hbar}[H_S,\rho(t)]\\\label{GMEfin}
&-&\frac{1}{\hbar^2}\sum_{l=L,R}\int dq\:\chi^l(t)
([{\cal T}_l,\Omega_{ql}(t)]+h.c)
\end{eqnarray}
where we have introduced the operators (see Ref. \onlinecite{Moldo2} for details):
%
%\begin{widetext}
\begin{eqnarray}\nonumber
&&\Omega_{ql}(t)=e^{-itH_S} \int_{t_0}^tds\:\chi^l(s)\Pi_{ql}(s)e^{i(s-t)\varepsilon^l(q)}e^{itH_S},\\\nonumber
&&\Pi_{ql}(s)=e^{isH_S}\left ({\cal T}_l^{\dagger}\rho(s)(1-f^l)-\rho(s){\cal T}_l^{\dagger}f^l\right )e^{-isH_S},\\\nonumber
&&{\cal T}_l(q)=\sum_{\alpha,\beta}{\cal T}_{\alpha\beta}^l(q)|{\bf \alpha}\rangle\langle {\bf \beta}|,\\
&&{\cal T}_{\alpha\beta}^l(q)=\sum_nT^l_{nq}\langle {\bf \alpha}|d_n^{\dagger}|{\bf \beta}\rangle.
\end{eqnarray}
%\end{widetext}
%
It is clear that ${\cal T}_{\alpha\beta}^l(q)$ describes the `absorption'
of electrons from the leads to the system and changes the many-body
states of the latter from $\beta$ to $\alpha$.  Observe that ${\cal
T}_{\alpha\beta}^l(q) \neq 0$ only if the number of electrons in
the many-body states $\alpha$ and $\beta$ differ by one. $f_l$
denotes the Fermi function in the lead $l$. The difference between
the chemical potentials defines the bias applied across the sample
$eV=\mu_L-\mu_R$. Observe also the presence of loss and gain terms
in $\Pi_{ql}$.

We also define a set of relevant states located in the energy window
$[E_{\rm min},E_{\rm max}]$, where $E_{\rm min}<\mu_R<\mu_L<E_{\rm max}$.
This 'active' window includes only those states of the sample that are
relevant to the transport.  More precisely, $E_{\rm min}$ is chosen such
that the levels with lower energy are fully occupied both prior to the
coupling of the leads, i.e. for $t \leq 0$, and also after the coupling
began, at $t > 0$, in the presence of the bias. Similarly, $E_{\rm max}$
is selected such that all states with higher energy are permanently empty.
Consequently the states outside this energy window do not contribute to
the current.  Based on this picture we conclude that it is sufficient
to compute an 'effective' reduced density matrix by taking into account
only those many-body configurations resulting from the single particle
states within the active window.  Also for the simplicity of notation
we shall specify in the many-body states only the occupation numbers
of the single-particle states within the active window.  Of course, the
validity of this truncation should be checked in the numerical simulations
by gradually enlarging the active window $[E_{\rm min},E_{\rm max}]$
until the calculated currents become stable.

The time evolution of the charge residing in the active region is related
to the diagonal elements of the reduced density matrix:
\begin{eqnarray}\nonumber
\langle Q_S(t)\rangle =\sum_n \sum_{\nu} i^{\nu}_n \, \langle\nu | \rho(t) | \nu\rangle .
\end{eqnarray} 
Using the GME, Eq.\ (\ref{GMEfin}), one can easily identify the contribution of
each level $n$ to the currents in the left and right lead:
\begin{eqnarray} \label{currents}
&& J_l=\sum_n J_{l,n} \nonumber \\
&& J_{l,n}= -\frac{1}{\hbar^2}\sum_{\nu}i^{\nu}_n\int dq\:\chi^l(t)
\langle \nu | [{\cal T}_l,\Omega_{ql}(t)]+h.c. | \nu \rangle  \nonumber \\
\end{eqnarray}
where $i_n^{\nu}=0,1$ is the occupation number of the $n$-th single
particle state inside the active window. The sign of the net currents in
the leads is positive if it is oriented from the left to the right, i.e.
$J_L>0$ if the electrons flow from the left lead towards the sample
and $J_R>0$ if they flow from the sample towards the right lead. During the
transient regime the sign of the net currents may change in time.

The GME is numerically solved through the Crank-Nicolson method (see
the details in Ref.\,(\onlinecite{Moldo2})). Throughout this work the Coulomb
interaction effects are not considered; further discussion on this point is given at the end
of Section III.

\section{Numerical results}

\begin{figure}[tbhp!]
\includegraphics[width=0.4\textwidth]{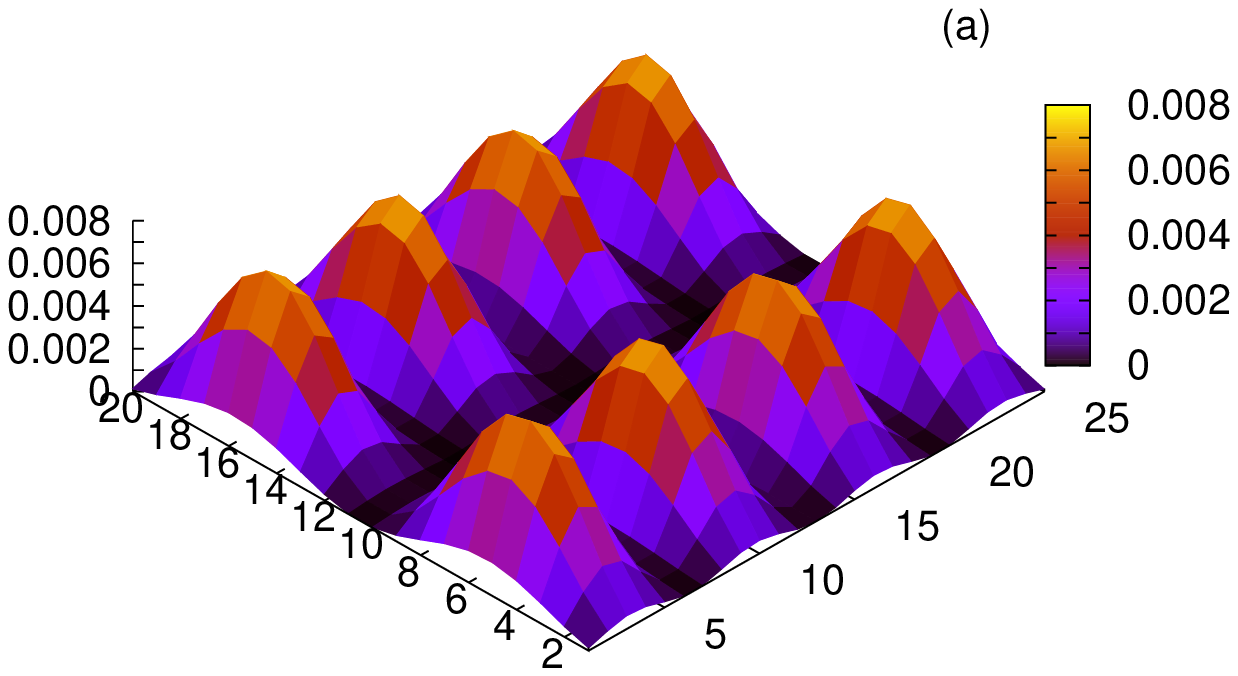}
\vskip -0.75cm
\includegraphics[width=0.4\textwidth]{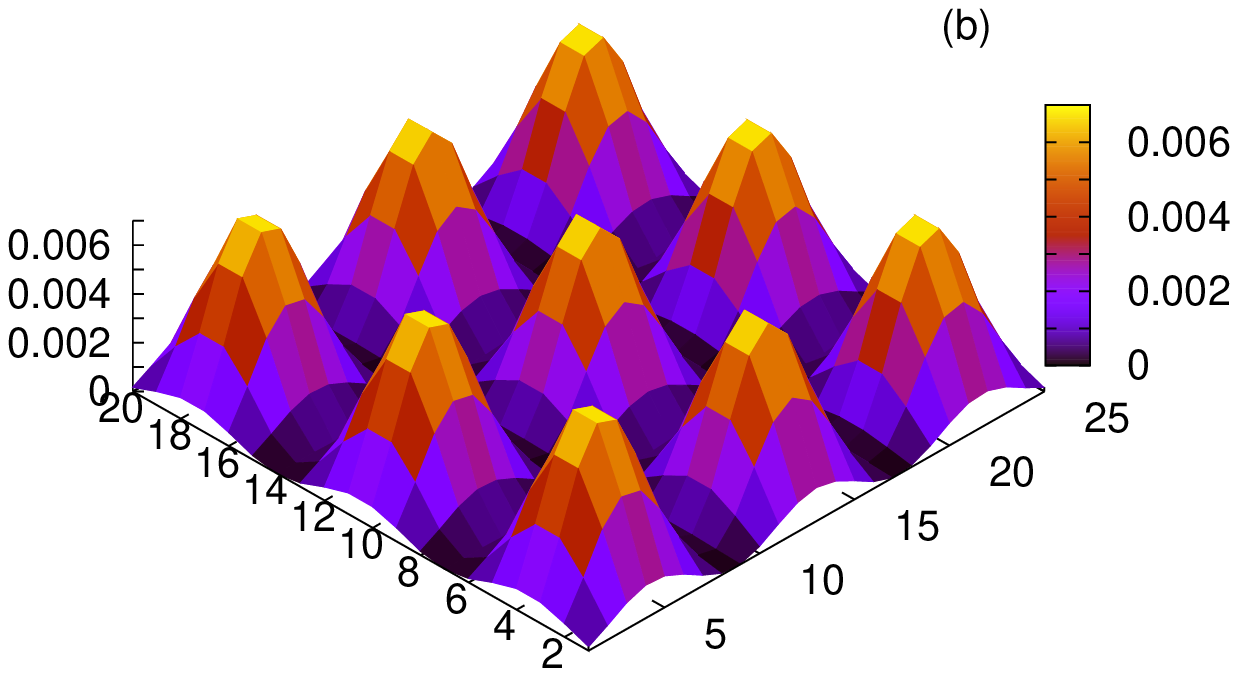}
\vskip -0.75cm
\includegraphics[width=0.4\textwidth]{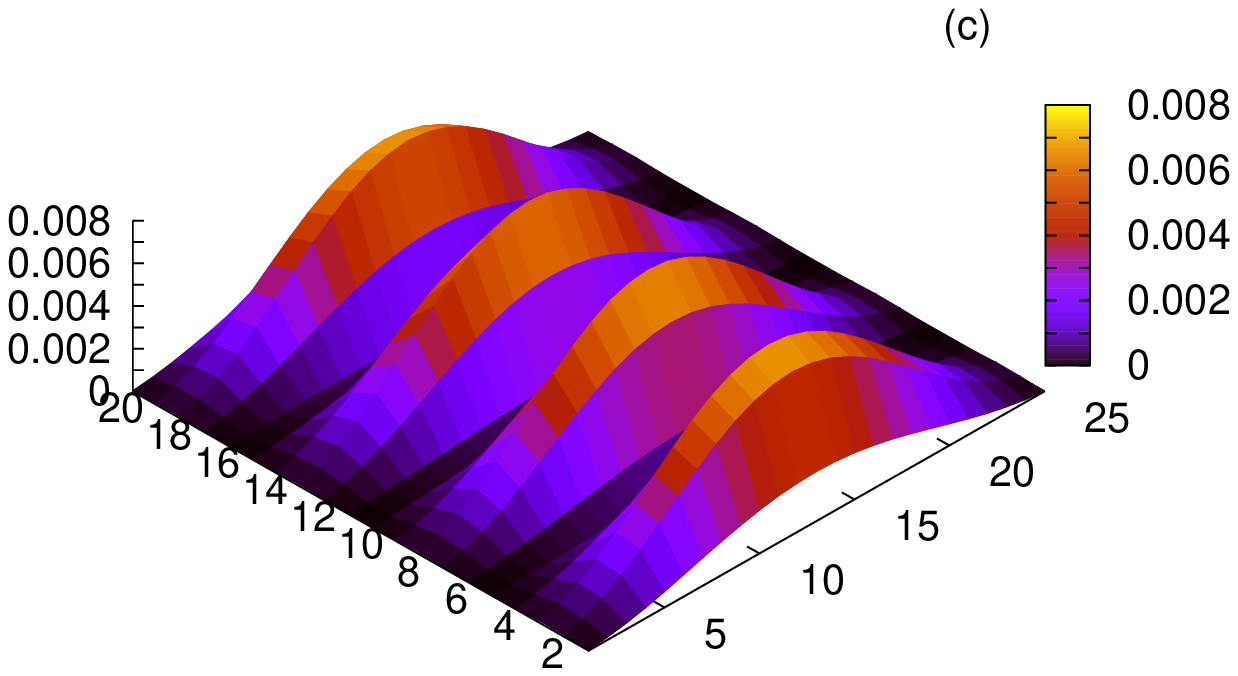}
\vskip -0.75cm
\includegraphics[width=0.4\textwidth]{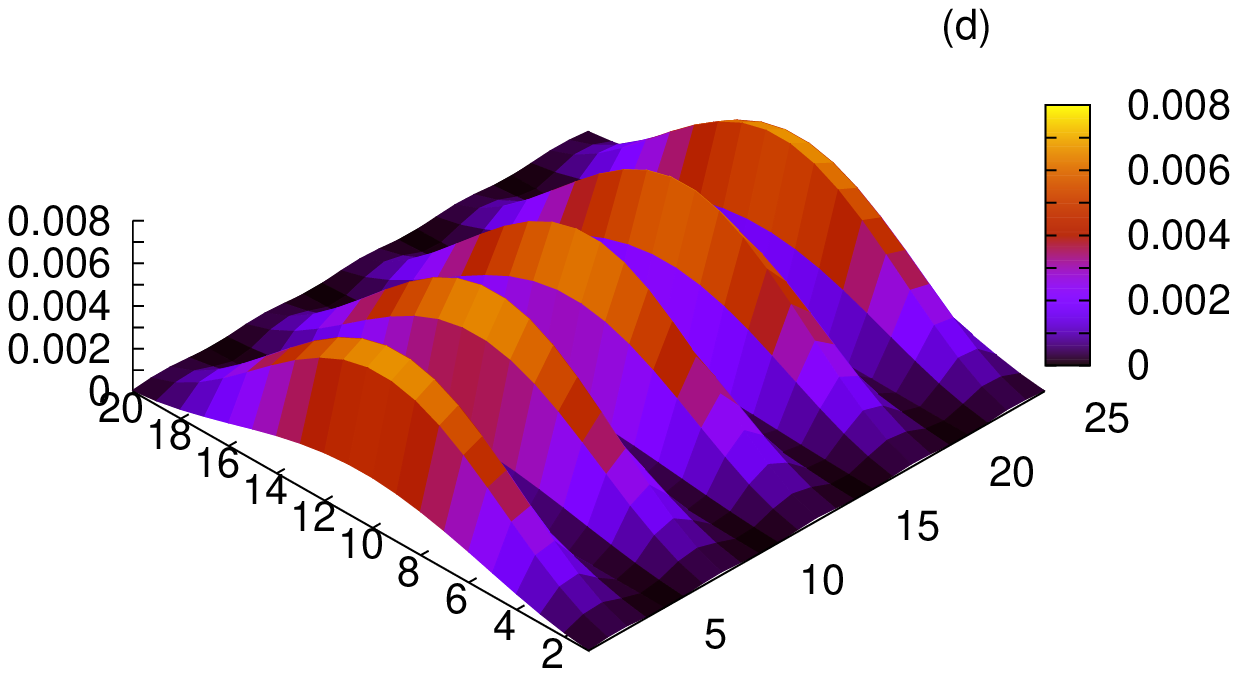}
\caption{(Color online) The localization probabilities associated
to the four states included in the active window. $E_{10}=-3.68$ $(k=1)$,
$E_{11}=-3.67$ $(k=2)$, $E_{12}=-3.63$ $(k=3)$ and $E_{13}=-3.62$ $(k=4)$ 
(in $t_D$ units).  The left lead is located near the lower left 
corner of the sample and the right lead coupled close to the upper 
right corner.}
\label{figure1}
\end{figure}

The first system we consider is a rectangular two-dimensional lattice
having $25\times 20$ sites.  The corresponding Hamiltonian has 500
eigenvalues $E_n$ and eigenvectors $\phi_n$, $n=1,2,...,500$, as many as
the number of sites.  The energy unit is given by the hopping parameter in
the sample $t_D=\hbar^2/2m^*a^2$, where $a$ is the lattice constant and
$m^*$ is the electron effective mass in GaAs. The spectrum of the isolated 
sample is contained in the range $(-2t_D,2t_D)$. For $a=8$ nm our lattice
describes a $200$ nm $\times 160$ nm sample. We fix $kT=0.1$ meV which
corresponds to a low temperature $T=1.15$ K. The sites of the system are
denoted by $i$ and are specified by the pair of coordinates $(i_x,i_y)$.
%The currents are expressed in $et_D/\hbar$ units.

In order to describe the gradual coupling of the leads to the sample and
the periodic modulation of the left contact we use specific coupling
functions $\chi_l(t)$.  The coupling to the leads begins at $t=0$
and evolves in time like $f(t)=1-2/(e^{\gamma t}+1)$.  The parameter
$\gamma$ defines the smoothness of the coupling. After some time $t_0>0$
the coupling to the left lead is turned off and on periodically. These
pulses are analytically defined by combining the functions $f(t)$ and
$1-f(t)$. We denote the pulse length by ${\tau}_p$.  We think this kind
of time-dependent perturbation is a reasonable model of an experiment
in which the sample is smoothly coupled to the leads with different
chemical potentials, possibly reaching an equilibrium state before the
pulses begin to act.

\begin{figure}[tbhp!]
\includegraphics[width=0.45\textwidth]{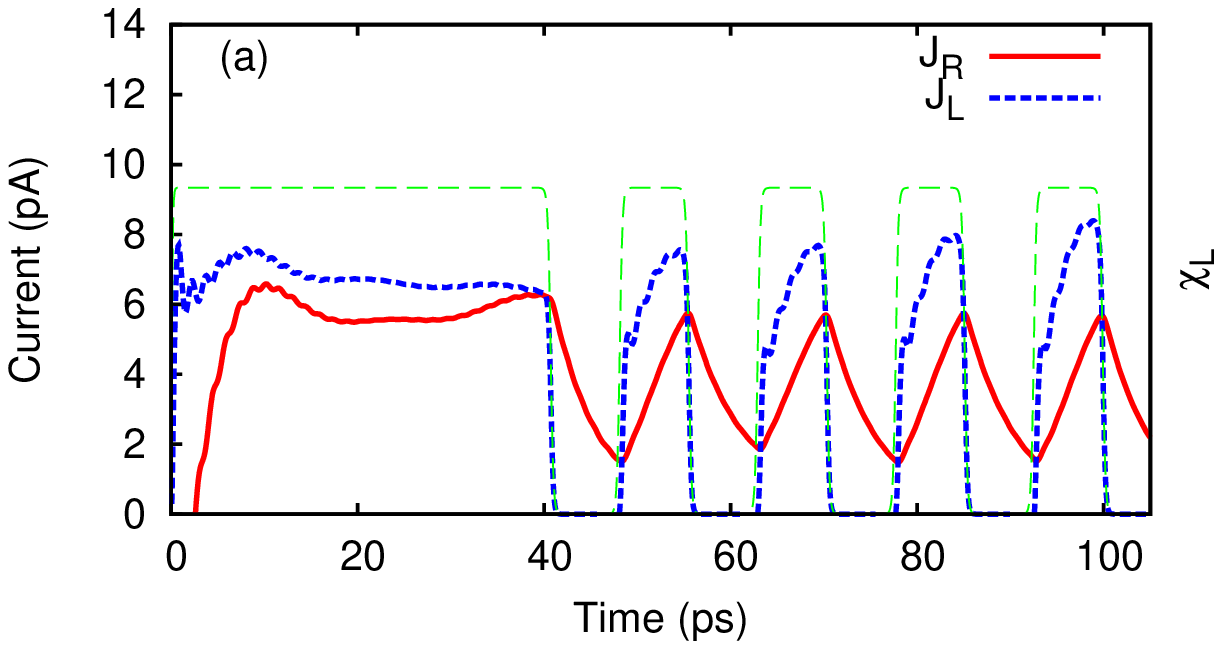}
\vskip -1.5cm
\includegraphics[width=0.45\textwidth]{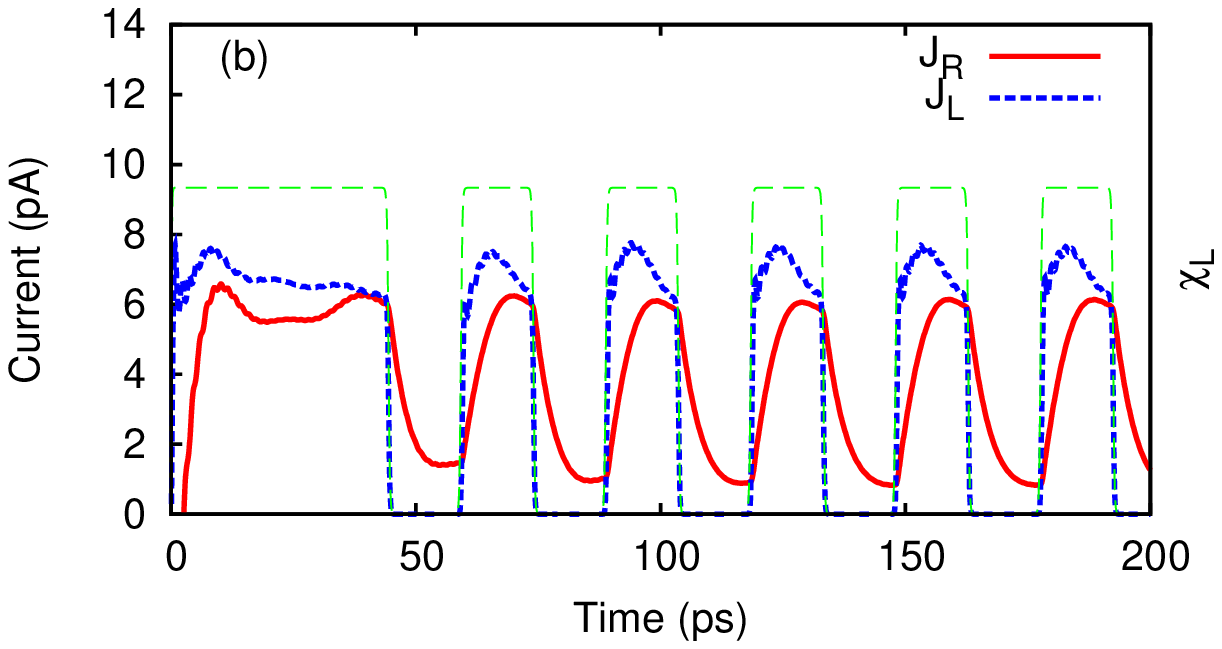}
\vskip -1.5cm
\includegraphics[width=0.45\textwidth]{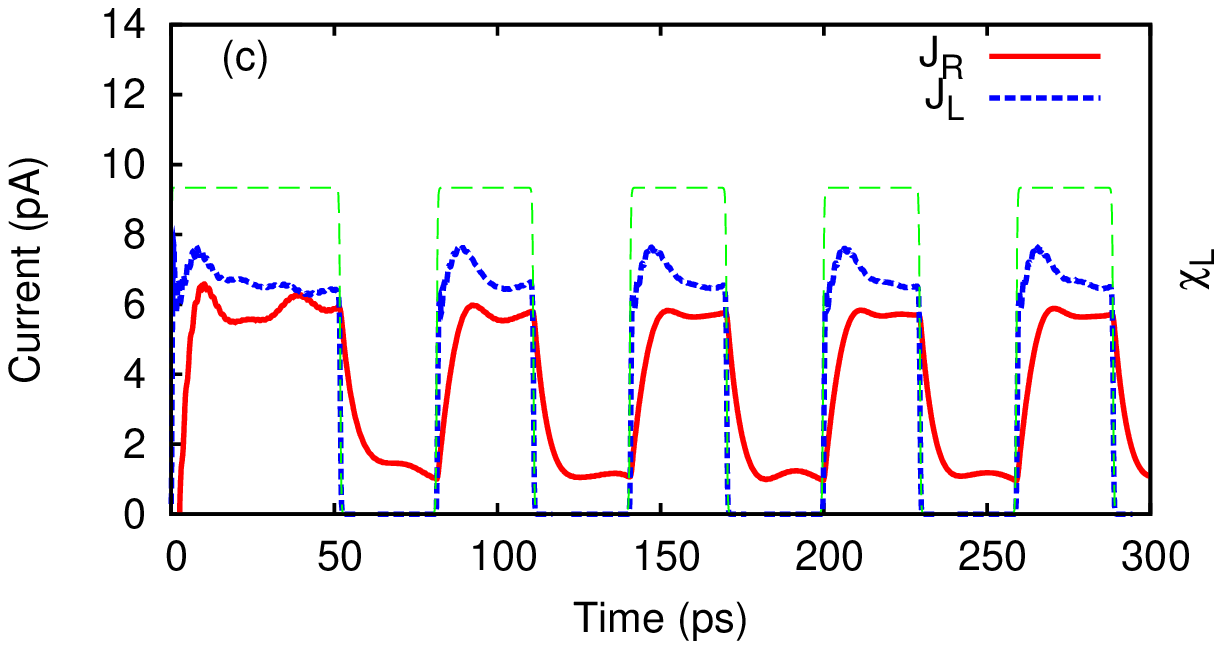}
\caption{(Color online) (a)-(c)Total currents in the input (left)
and output (right) leads for the $25\times 20$ sites system for
different pulse lengths: (a) ${\tau}_p=7.5$ ps, (b) ${\tau}_p=15$ ps,
(c) ${\tau}_p=30$ ps.  The bias window covers only two single particle
states. In Figs. (b) and (c) we show the currents for longer times
than in figure (a) in order to capture the same number of pulses.
The signal applied on the input contact is also qualitatively shown 
with a dashed line.
The coupling strength $V_{L}=V_{R}=4$, $\mu_L=-3.65$ and $\mu_R=-3.7$.}
\label{figure2} 
\end{figure}
\begin{figure}[tbhp!]\includegraphics[width=0.45\textwidth]{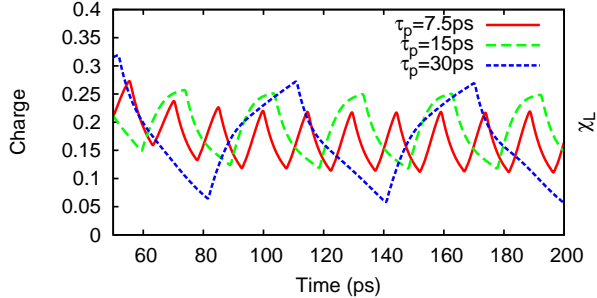}
\caption{(Color online) The total charge accumulated on the two states 
from the bias window for the three pulse lengths considered in Figs.\,2(a)-(c). }
\label{figure3}
\end{figure}

The currents depend strongly on the placement of the contacts.  In this
example the two leads are attached at diagonally opposite corners of
the sample.  We first choose the chemical potentials $\mu_L=-3.65$  and
$\mu_R=-3.60$. Then our sample has nine energy levels below $\mu_R$,
and as we checked they do not contribute significantly to the transport.
The chosen active region contains two states within the bias window
which are $E_{10}$ and $E_{11}$ and two more states above $\mu_L$,
which are $E_{12}$ and $E_{13}$.  Instead of the label $n=10,11,12,13$
for the states in the active window (in the present example), we will
also use the label $k=1,2,3,4$ respectively.

In Figs.\,1(a)-(d) we show the site occupation
probabilities $|\phi_n(i)|^2$ for the active states.  We emphasize that
the middle state with $n=12$ or $k=3$, Fig.\,1(c), is weakly coupled to the
leads while the other ones have a larger, but still moderate coupling.  

Figs.\,2(a)-(c) show the time-dependent total currents in both leads
for different pulse lengths.  We also indicate qualitatively the pulsed
signal as given by the function $\chi_L(t)$ (arbitrary units are used on
the corresponding axis).  In the beginning the system is coupled to the
leads and both contacts are kept open for a time $t_0$.  Then the the left
contact is modulated by pulses while the other one is left open. As the
sample dynamics adapts to the input signal a periodic regime is already
established after two pulses.  The first observation is that for very
short pulses ($\tau _p=7.5$ ps in Fig.\,2(a)) the output current $J_R$
has a triangular shape although the modulating signal is rectangular.
In addition $J_R$ does not vanish when the left lead is disconnected,
resembling the charging-relaxation characteristic of a capacitor.
The charge is first absorbed from the left lead when the contact is
switched on, and then only partially expelled into the right lead,
when the contact to the left lead is off, i.e. even in the absence of
a driving bias. The decay of the output signal looks exponential, but
after one complete cycle, i.e. before the left contact opens again, the
magnitude of the output current in the right lead is still considerable.
The current in the left lead $J_L$ decays much faster when the contact
is turned off, following closer the modulation potential at the contacts.
We notice that right after the left contact opens the current is injected
in the sample quite fast but the current in the right lead increases
slower.

When the pulse length increases the shape and the amplitude of the output
current change considerably.  $J_R$ reaches maxima even before the pulse
is turned off and remains almost constant during the second half of the pulse,
while at the same time the input current decrease. However, at $\tau_p=15$ps
the output current still shows an exponential decay and does not reproduce
the input signal.
\begin{figure}[tbhp!]
\includegraphics[width=0.45\textwidth]{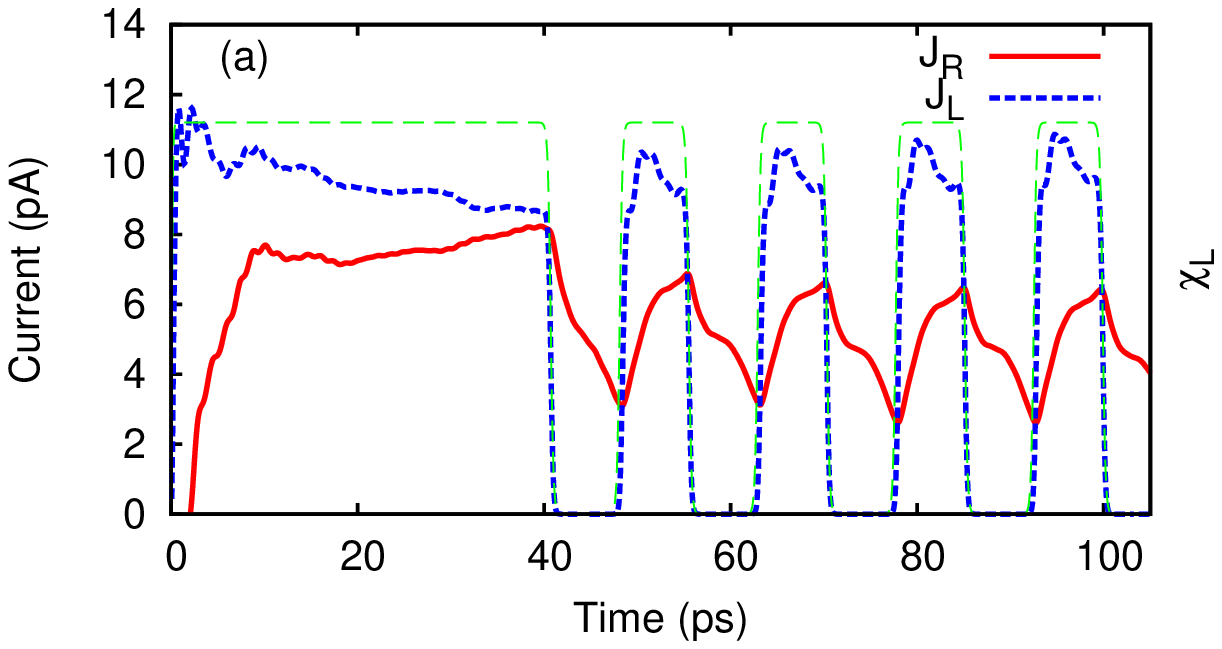}
\vskip -1.5cm
\includegraphics[width=0.45\textwidth]{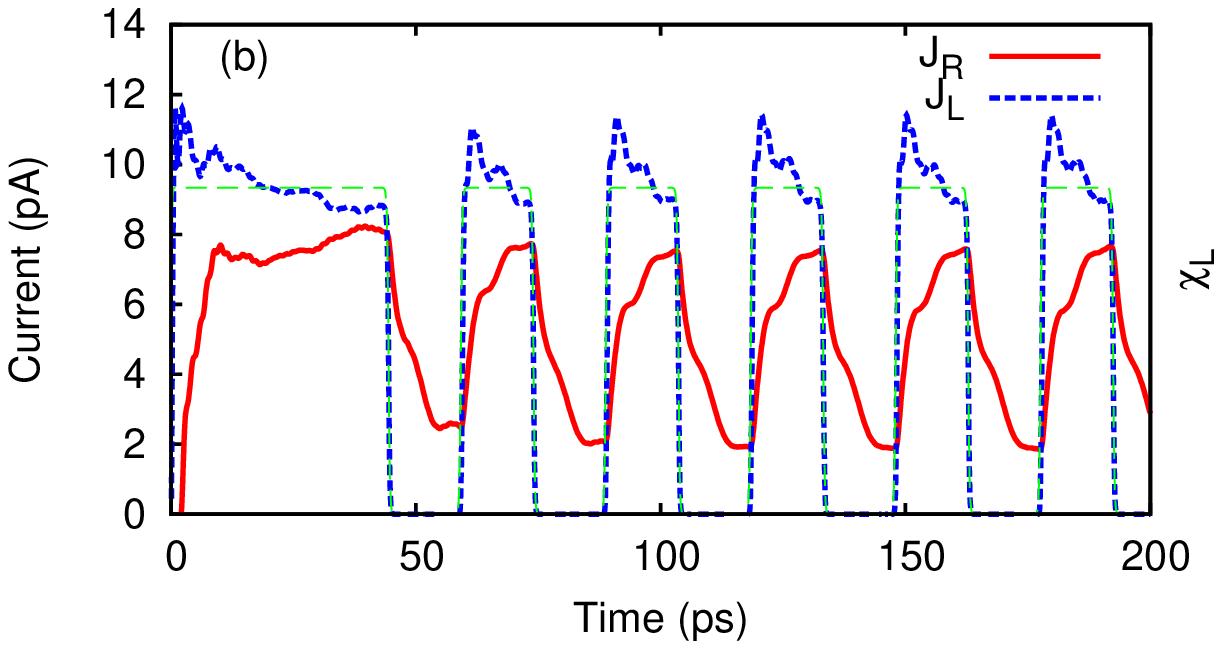}
\vskip -1.5cm
\includegraphics[width=0.45\textwidth]{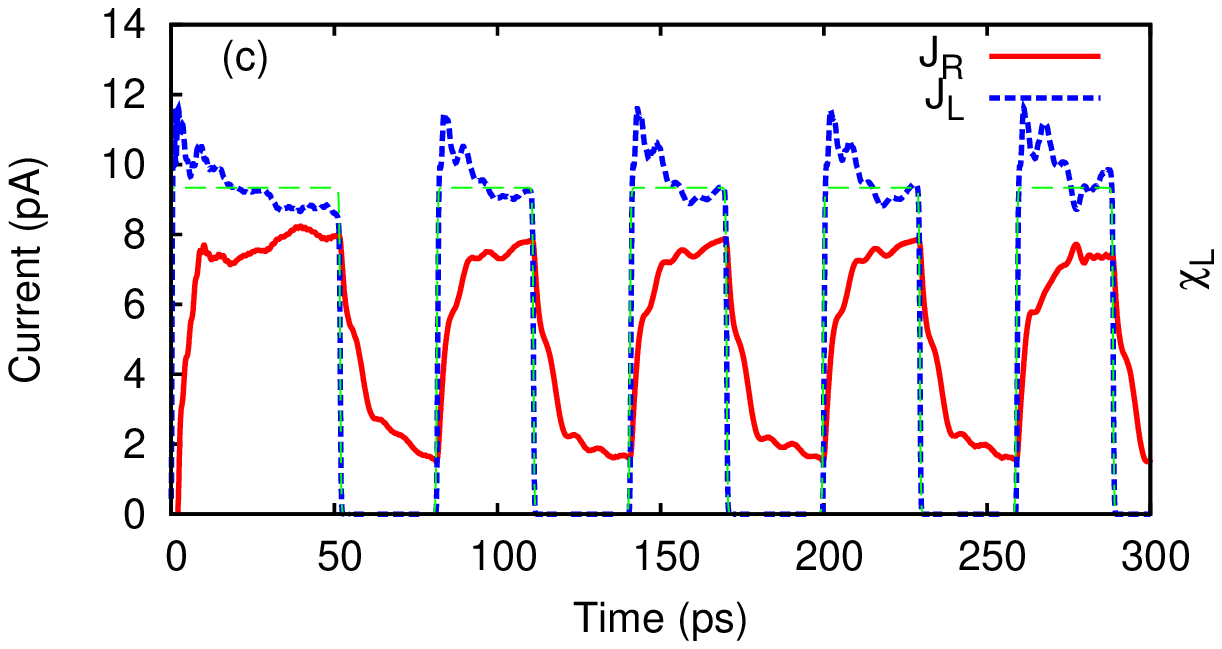}
\caption{(Color online) (a)-(c)Total currents in the input (left)
and output (right) leads for the $25\times 20$ sites system for
different pulse lengths: (a) $\tau _p=7.5$ ps, (b) $\tau _p=15$ ps, (c)
$\tau_p=30$ ps. The bias window covers four single particle states. The
signal applied on the input contact is also given in arbitrary units-
dashed line. The coupling strength $V_{L}=V_{R}=4$.  $\mu_L=-3.6$
and $\mu_R=-3.7$.}
\label{figure4}
\end{figure}
A different response to the pulse train is obtained at $\tau_p=30$ps. In
this case the output current decays first exponentially and then stays
flat until the left contact opens again.

Comparing the behavior of the input currents in all three cases we see
that in the first half of the 30 ps pulse the current is similar to the
one that develops in the full length of the 15 ps pulse. The saw-tooth
profile of the current in Fig.\,2(a) is also present in the first half
of the 15 ps pulse. This is due to the fact that as long as the sample
is coupled to both leads it has the transient behavior which is
already observed during the initial charging time. It is also clear that
if the pulse length is too short the left lead does not feed enough
charge to the sample in order to maintain a constant output current.

\begin{figure}[tbhp!]
\includegraphics[width=0.45\textwidth]{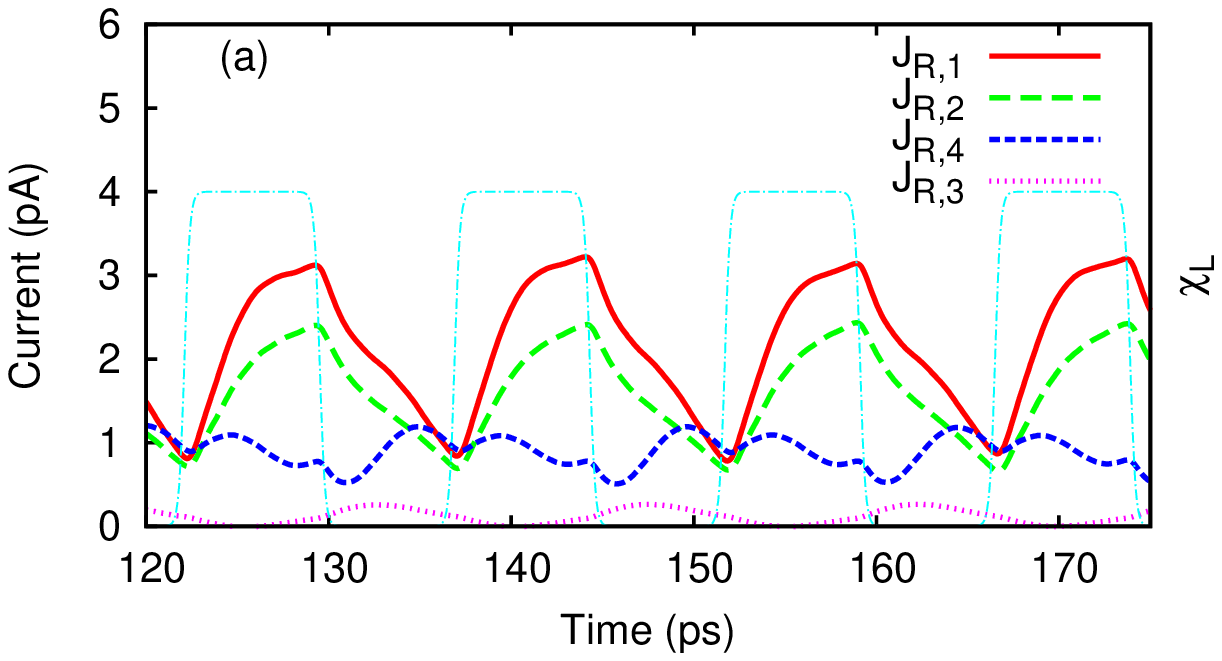}
\vskip -1.5cm
\includegraphics[width=0.45\textwidth]{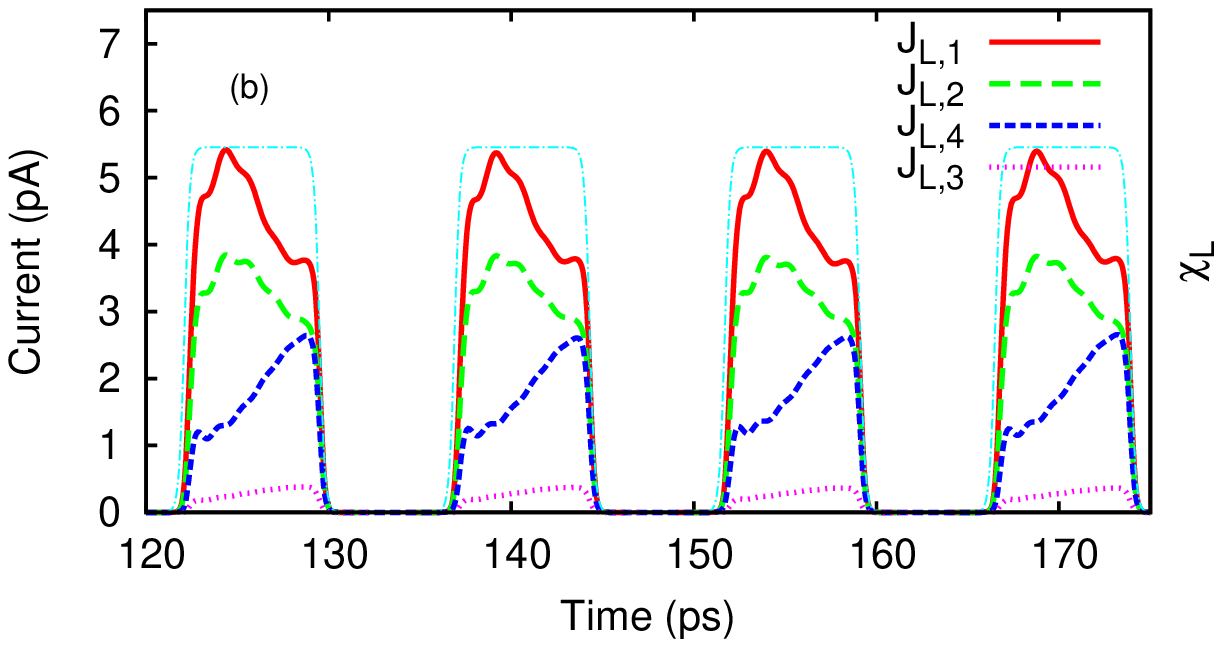}
\vskip -1.5cm
\includegraphics[width=0.45\textwidth]{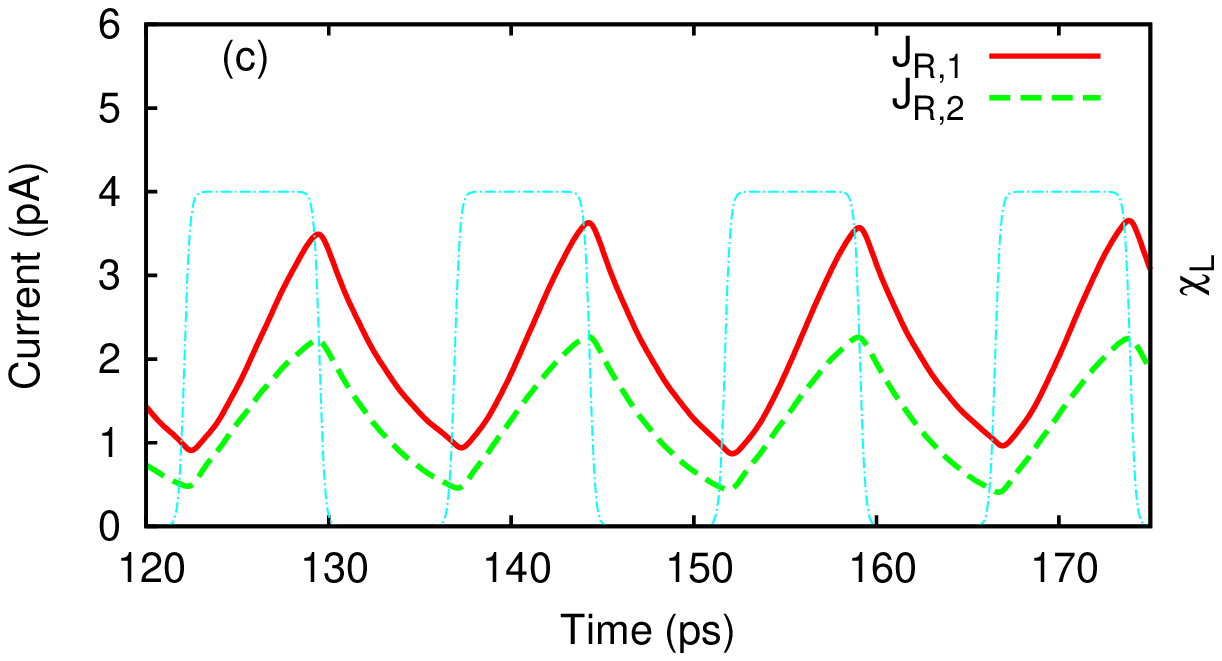}
\caption{(Color online) Partial currents in the output (a) and input
(b) leads for the $25\times 20$ sites system for $\tau _p=7.5$
ps and $\mu_L=-3.6$ and $\mu_R=-3.7$.  In this case the bias window
covers four states.  (c) The output partial currents for the two-level
configuration $\mu_L=-3.65$ and $\mu_R=-3.7$.  The coupling strength
is $V_{L}=V_{R}=4$.}
\label{figure5}
\end{figure}

An estimate of the pulse length which generates an output current
with almost a rectangular shape can be taken from the first transient period
($t<t_0$); that pulse length should be at least equal to the
time at which the output current become nearly equal.

In Fig.\,3 we compare the total charge accumated on the two states within the
bias window for the three pulse lengths. As the pulse length increases
more charge is transferred through the system and therefore the output
current increase. One should notice that for the 30 ps pulse the charge
first relaxes exponentially but then almost linearly. Since the current
is essentially the derivative of the charge with respect to time this means
the current in the right lead is first exponential and then constant,
as we have already learned from Fig. \,2(c).

\begin{figure}[tbhp!]

\includegraphics[width=0.45\textwidth]{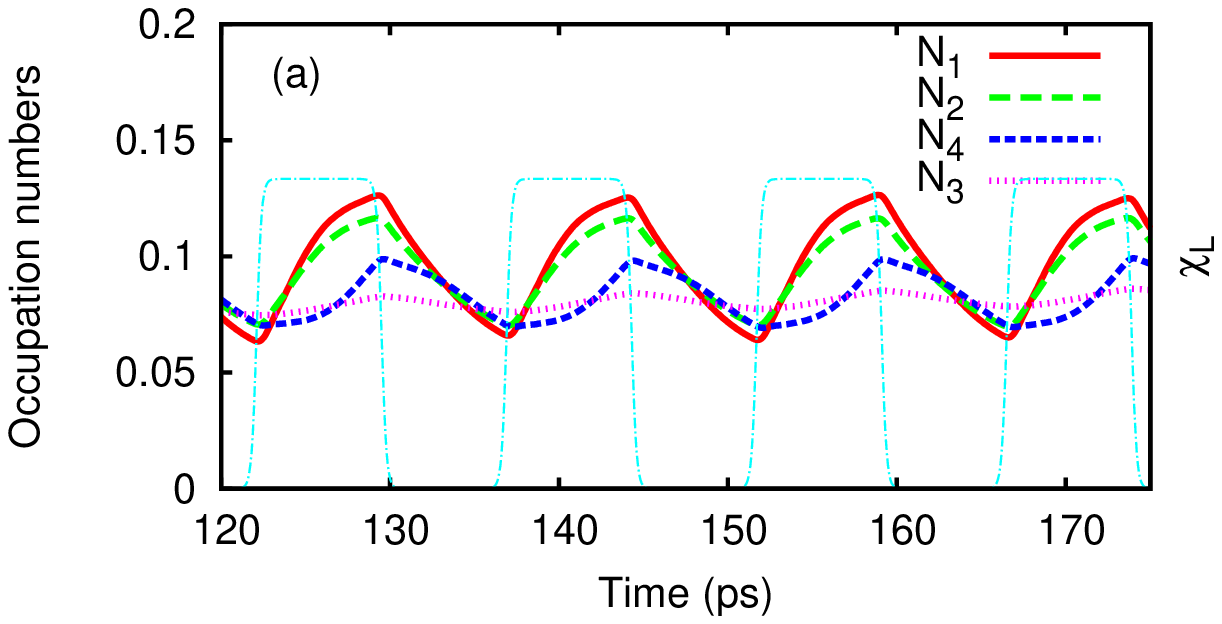}
\vskip -1.5cm
\includegraphics[width=0.45\textwidth]{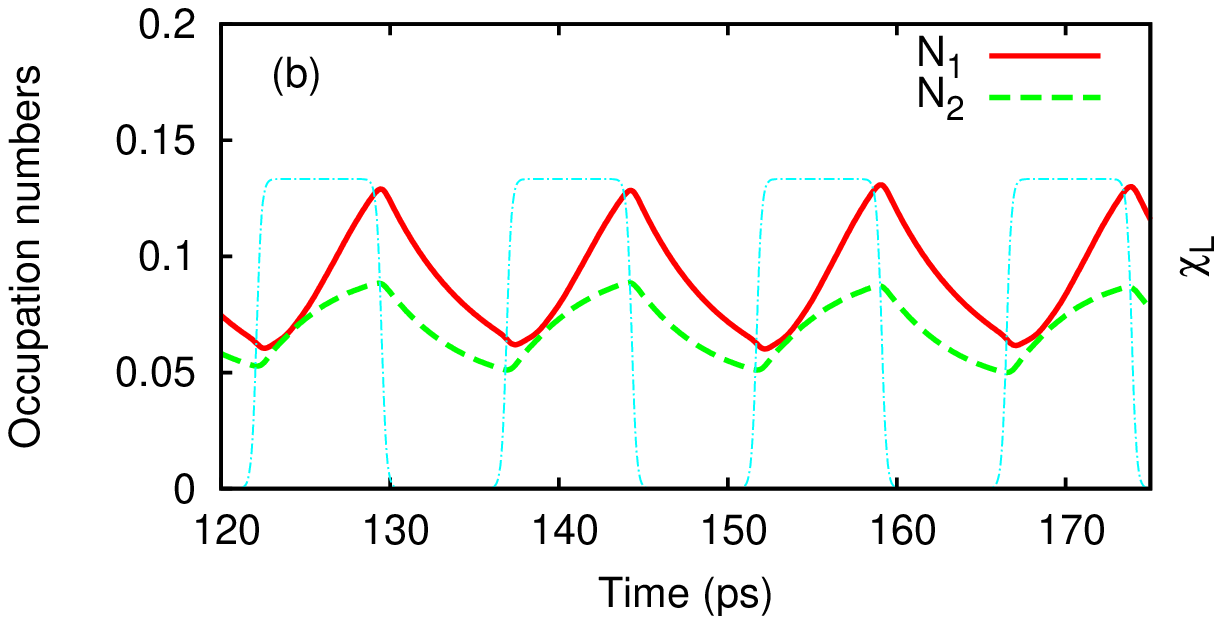}
\caption{(Color online) (a) The occupation numbers of states from the
bias window for a pulse length $\tau_p=7.5$ ps, $\mu_L=-3.6$ and $\mu_R=-3.7$. 
(b) The same in the two level case $\mu_L=-3.65$ and $\mu_R=-3.7$. }
\label{figure6}
\end{figure}

In Figs.\,4(a)-(c) we show the currents obtained from the same sample when
the chemical potential of the left lead is pushed to $\mu_L=-3.6$ and thus
two more states enter into the bias window, i.e. those with $k=3$ and
and $k=4$.  When comparing to Fig.\,2 we observe additional 'shoulders'
developing in the output current.  These shoulders are produced by the
new states included in the bias window during the relaxation into the
right lead. The relaxation processes depend on the coupling between the
states and the lead.  Since one of the two new states is only poorly
coupled to the leads, i.e.\ the one withe $k=3$ (see Fig.\,1(c)), the
shoulder is actually produced by the new state with $k=4$.

In the following we compare the partial currents in the right lead for the
two states and four states configurations in the case of the ultrashort
7.5 ps pulse.  The partial currents $J_{l,k}$ are calculated with
Eq.(\ref{currents}).  Fig.\,5(a) gives the output currents and confirms
that the 3rd state does not contribute significantly to the transport.
Both $J_{R,1}$ and $J_{R,2}$ increase during the entire pulse length,
although with a lower slope in the second part.  Rather surprisingly,
in this time interval $J_{R,4}$ decreases.  In the relaxation interval
the situation is the opposite: $J_{R,1}$ and $J_{R,2}$ are monotonously
decreasing while the current of the 4th state considerably increases in
the second half of the relaxation interval.  The currents entering the
sample from the left lead (see Fig.\,5(b)) also have interesting features:
$J_{L,1}$ and $J_{L,2}$ rise suddenly to a maximum value, while $J_{L,4}$
increases slower and does not reach a maximum within the pulse. This
suggests that the two lowest states absorb quickly more charge from the
left reservoir. By looking at the occupation numbers shown in Fig.\,6(a)
one convinces himself that this is indeed the case.  The occupation of
the 4th level increases much slower than $N_1$ and $N_2$.

The behavior of $J_{L,4}$  and $J_{R,4}$ can be explained by the dynamic
tunneling processes described by the gain and loss terms of the solution
of the master equation via the matrix $\Pi_{ql}$. Suppose, for example, that
at time $s<t$ an electron tunnels from the input lead into the lowest
state of the bias window, $k=1$ .  At instant $t$ the same electron
may tunnel out into the right lead.  But it is also possible that this
electron remains in the sample, while another electron, from another
state, say the 4th, tunnels back into the left lead; escaping into
the left lead from the 4th state is more likely than in the right lead
because the chemical potential $\mu_L$ is much closer to $E_{13}$ ($k=4$)
than $\mu_R$. Repeated tunneling from highest state of the sample towards
different leads and back into the sample lead to the charging of the lower
states at the expense of the higher ones.  Consequently the output currents
associated to the lower states 1 and 2 are increasing, see Fig.\,5(a),
whereas the output current $J_{R,4}$ decreases.  Note also that the
net input current $J_{L,4}$ is smaller than $J_{L,1}$ and $J_{L,2}$
because there are more electrons tunneling out from the level $k=4$.

In contrast, when the pulse is turned off, the charge on the 4th level
leaves the sample only via the output lead and therefore $J_{R,4}$
increase as shown in Fig.\,5(a). On the other hand the tunneling processes
from the left lead to the lowest levels are switched off and so $J_{R,1}$
and $J_{R,2}$ decrease on the relaxation interval.

For comparison we also show in Fig.\,5(c) and Fig.\,6(b) the output
currents for the two-level configuration ($k=1,2$) and the occupation
numbers of the two levels.  For the two-level configuration the
partial occupation numbers display similar charging/relaxation shapes.
This happens because these two particular states are equally coupled
to the leads and hence the input signal propagates almost identically
through both states towards the right lead.  Remarkably, a similar
behavior of the time-dependent occupation number was obtained from the
experimental data by Lai {\it et al.} \cite{Lai}

In order to get more informations on the relevant tunneling processes
we have analyzed the diagonal elements (i.e. the populations) of the
reduced density operator.  It useful to introduce a shorter notation
for the many-body states by interpreting the occupation numbers as
decimal numbers written as binary strings (but reading them in the
reverse order, and adding 1). For example $|1\rangle = |0000\rangle$, 
$|2\rangle = |1000\rangle$, $|3\rangle = |0100\rangle$ etc.

\begin{figure}[tbhp!]
\includegraphics[width=0.45\textwidth]{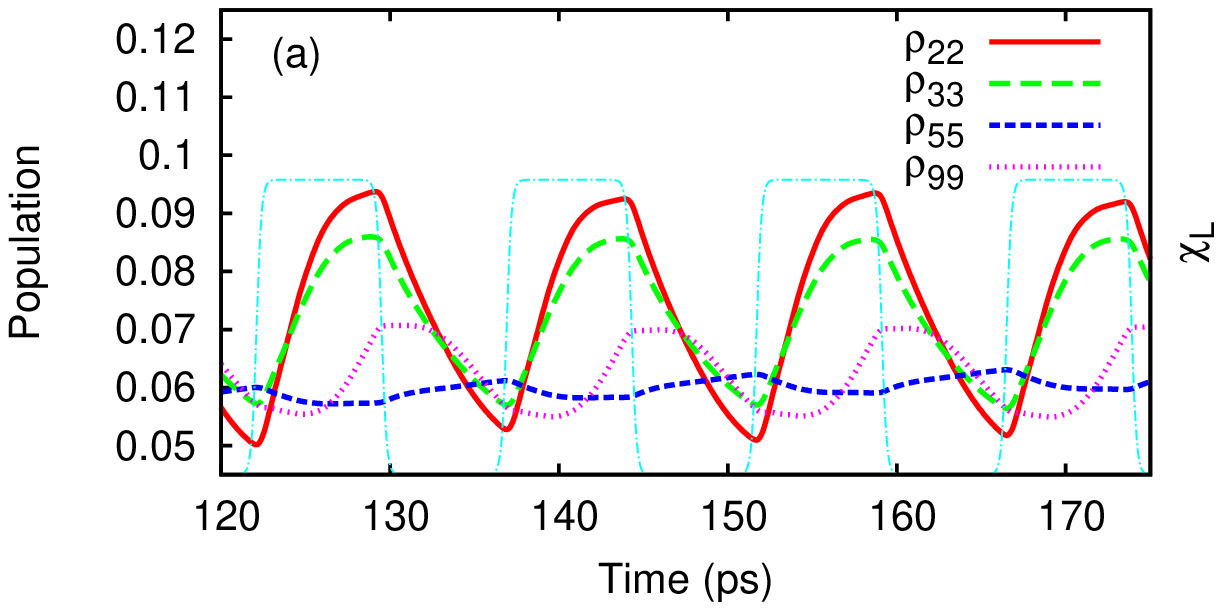}
\vskip -1.5cm
\includegraphics[width=0.45\textwidth]{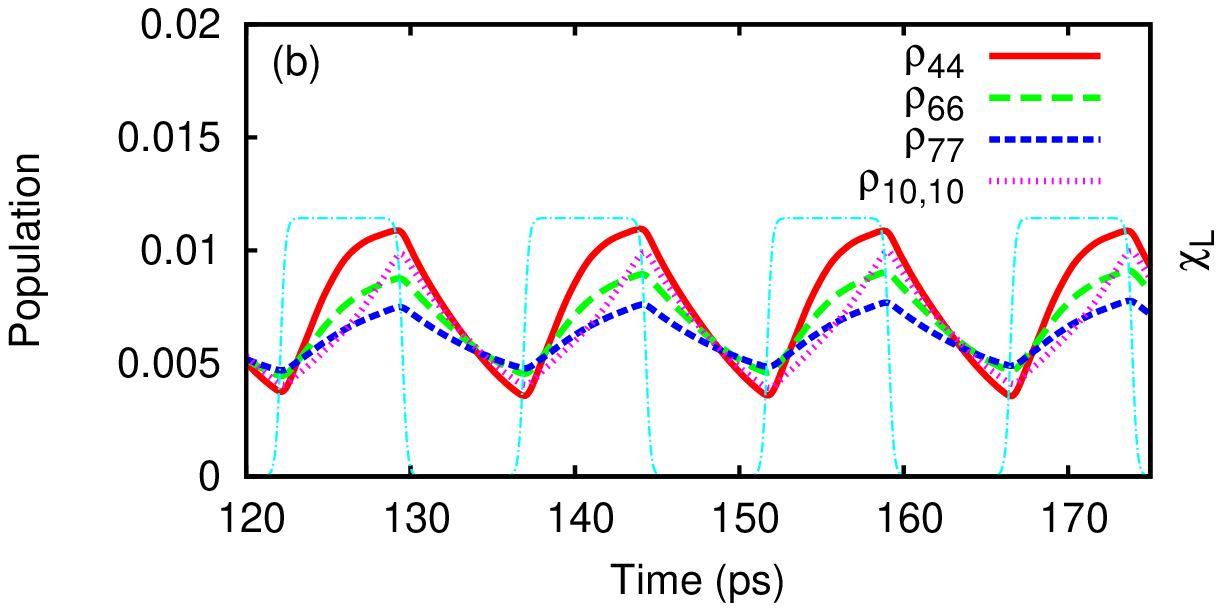}
\caption{(Color online) The relevant diagonal elements of the reduced
density operator for the $25\times 20$ sites system calculated with
four single particle states in the active region.  (a) The diagonal
elements of the reduced density operator corresponding to single-particle
configurations $|2\rangle = |1000\rangle$, $|3\rangle = |0100\rangle$,
$|5\rangle = |0010\rangle$, $|9\rangle = |0001\rangle$.  (b) The most
relevant two-particle configurations $|4\rangle = |1100\rangle$,
$|6\rangle = |1010\rangle$, $|7\rangle = |0110\rangle$ $|10\rangle =
|1001\rangle$.  $\mu_L=-3.6$ and $\mu_L=-3.7$.  }
\label{figure7}
\end{figure}

Fig.\,7(a) shows the populations corresponding to the single-particle
sector of the Fock space, that is, the probabilities for all
the configurations containing only one electron. We also give in
Fig.\,7(b) the most relevant two-particle configurations. The vacuum
state $|0000\rangle$ has the largest probability $\rho_{11}$ which is
not shown.  At $t=0$ it is 1 because the sample is initially empty, then
during the charging period it drops to about 0.6, but it increases back
to about 0.7 during the relaxation interval. 
From Fig. \,7(a) we infer that i) the configurations $|1000\rangle$
and $|0100\rangle$ are the most probable in the transient regime;
ii) the corresponding probabilities $\rho_{22}$ and $\rho_{33}$
decay exponentially (qualitatively speaking) during the relaxation,
but $\rho_{99}$ which is the probability of the state $|0001\rangle$
does not decay exponentially.  This state is even stable for some
time in the relaxation regime, while the probabilities for the states
$|1000\rangle$ and $|0100\rangle$ decrease.  This is another way of
seeing that the 4th state relaxes later into the right lead.  iii)
$\rho_{55}$ is only slowly increasing and has only small oscillations
due to the time-dependent signal, due to the weak coupling to the leads.
The two-particle configurations shown in Fig.\,7(b) have very all small
probability because the system is poorly charged during the ultra-short
pulse considered here.

\begin{figure}[tbhp!]
\includegraphics[width=0.45\textwidth]{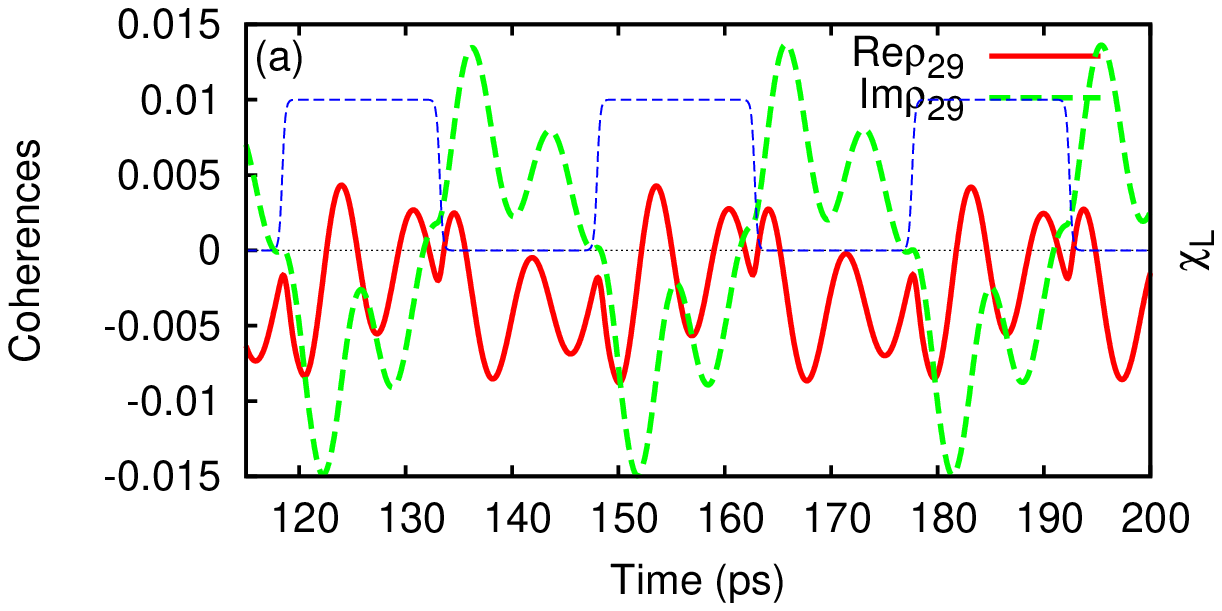}
\vskip -1.5cm
\includegraphics[width=0.45\textwidth]{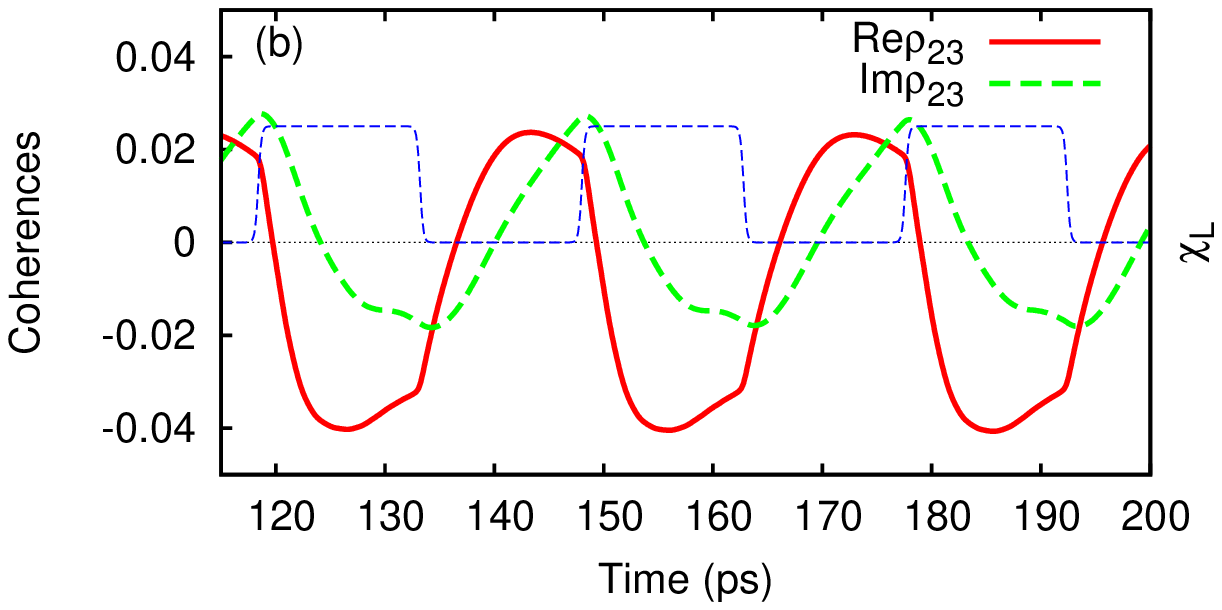}
\vskip -1.5cm
%\includegraphics[width=0.45\textwidth]{c59.eps}
%\vskip -1.5cm
\includegraphics[width=0.45\textwidth]{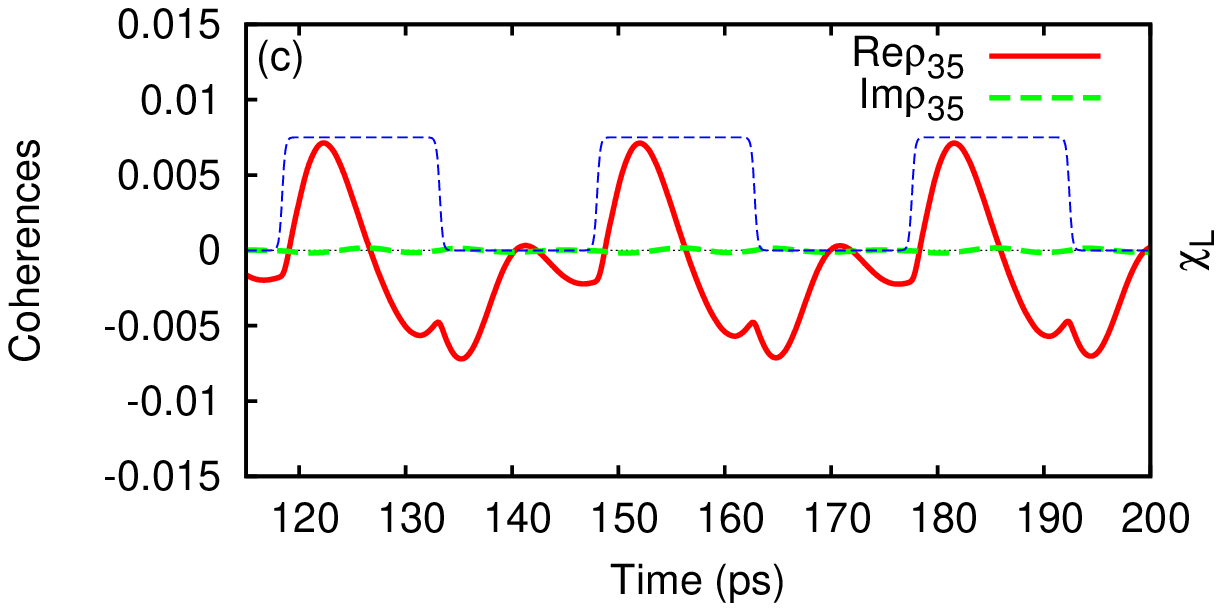}
\caption{(Color online) (a)-(c) The imaginary and real parts of the 
off-diagonal elements of the reduced density matrix which couple various 
states (see the discussion in the text). Note the multiple oscillations 
in any half-period and the small amplitude for $\rho_{29}$.  The dashed (blue) line represents the signal applied on the input contact. $\tau_p=15$ ps, 
$\mu_L=-3.6$ and $\mu_L=-3.7$.}
\label{figure8}
\end{figure}

\begin{figure}[tbhp!]
\includegraphics[width=0.45\textwidth]{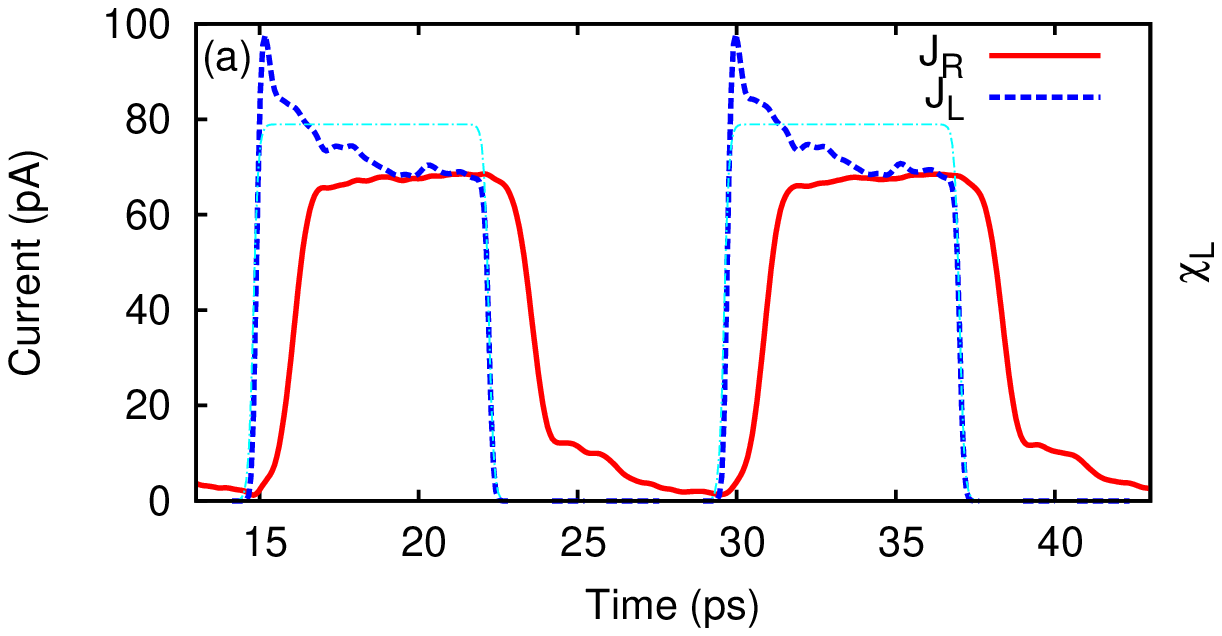}
\vskip -1.5cm
\includegraphics[width=0.45\textwidth]{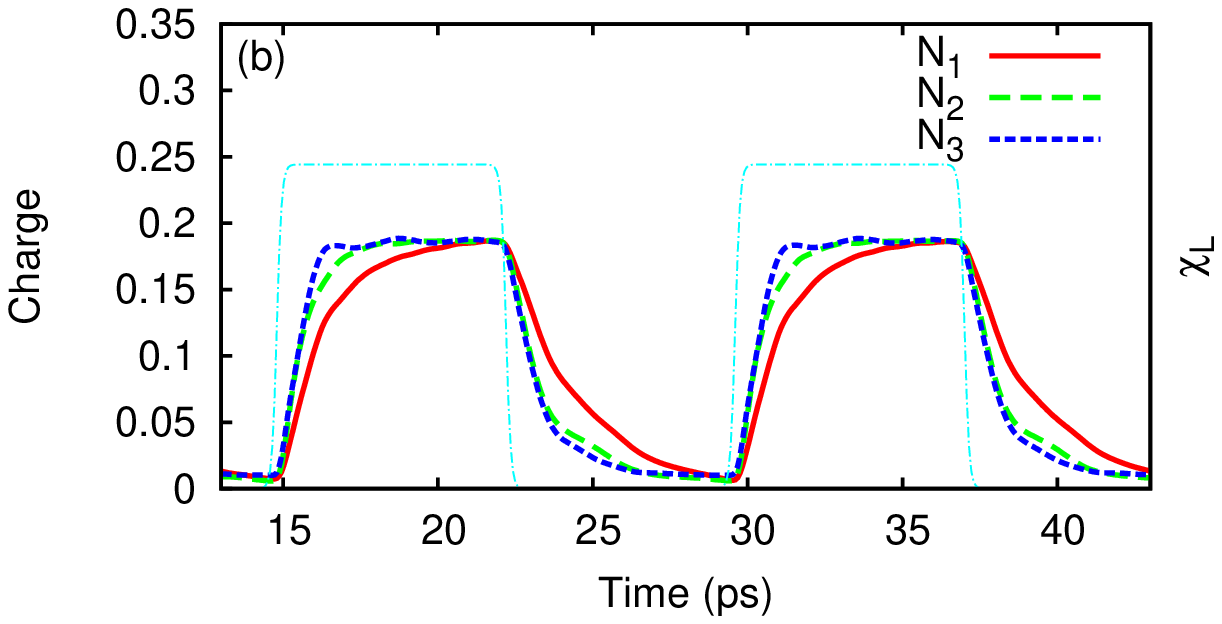}
\vskip -1.5cm
\includegraphics[width=0.45\textwidth]{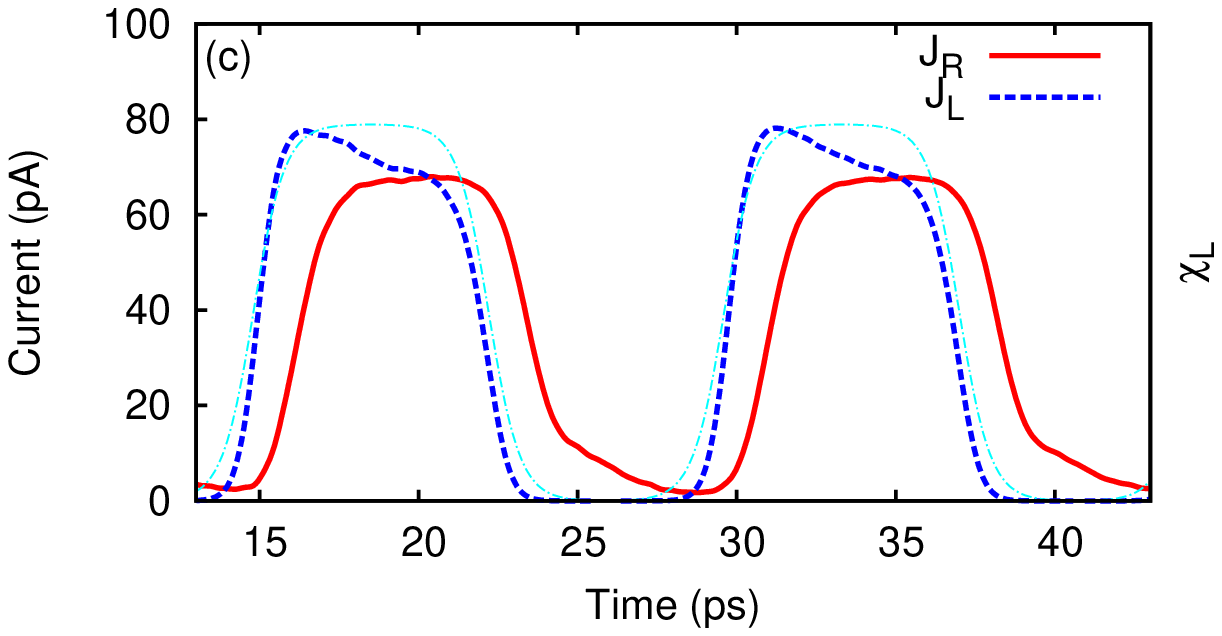}
\caption{(Color online) (a) The total input and output currents  for a
$5\times20$ sites system with three states within the bias window. $J_R$
is delayed w.r.t. to $J_L$. The pulse is rather squared (the dashed
(green) line), with the rise time defined by the parameter $\gamma=1$. (b)
The occupation numbers of the states within the bias window. Note that
the lowest states relaxes at a slower rate due to a smaller coupling to
the leads. (c) The total input and output currents for a smooth pulse
($\gamma=0.25$).  $\tau_p=15$ ps,$\mu_L=-3$ and $\mu_L=-3.6$.}
\label{figure9}
\end{figure}

For completeness we give in Figs.\,8(a)-(c) the off-diagonal elements
of the reduced density matrix, known as "coherences", which are related
to the transitions between different states. For example $\rho_{29}$
(Fig.\,8(a)) describes the process in which electrons can tunnel from
the 4th state into the leads and then back to the sample on the 1st
state. We see that the real and imaginary parts of the coherences have
oscillations and change sign.  Moreover, the oscillations are more
complex if the matrix element implies single particle states separated
by other energy levels. Indeed, by comparing ${\rm Re}\rho_{23}$ and
${\rm Re}\rho_{29}$ we see that the latter displays two minima and
two maxima in the charging time interval, while ${\rm Re}\rho_{23}$
has only one minimum.  This feature suggests again that electrons
make transitions via intermediate single-particle states as follows:
initial state $\to$ intermediate state $\to$ leads $\to$ final state.
We have checked that coherences between states with only one intermediate
state in between (e.g.\ ${\rm Re}\rho_{39}$) have only one maximum and 
one minimum.

Another remark is that the oscillations of $\rho_{23}$ have a higher
amplitude than the other two coherences, which is consistent to the fact
that the two configurations $|2\rangle = |1000\rangle$ and $|3\rangle
= |0100\rangle$ are the most probable ones in the stationary regime.
Fig.\,8(c) shows that the 3rd level is very weakly coupled to the
level below, which is expected since that state is poorly coupled to the
leads. By increasing the pulse width the number of oscillations during the
pulse generally increases as more and more transitions may occur. Also,
two-particle configurations will develop, while the probability of the
single-particle states decrease (not shown).  It should be mentioned
here that in the non-Markovian approach the coherences also contribute to
the currents because they are coupled to the diagonal elements of $\rho$.

We continue with similar results for a different sample, a narrow 40
nm $\times$ 160 nm quantum wire described by a $5\times 20 = 100$ sites
lattice. The system has now 100 eigenstates and it is connected to two
three-channel leads on both narrow sides. For simplicity we consider
uncoupled channels, which means the electrons cannot jump between
the channels.  The chemical potentials of the leads are $\mu_L=-3$
and $\mu_R=-3.6$; they cover three states now, $E_3=-3.53$, $E_4=-3.38$
and $E_5=-3.19$ (or $k=1,2,3$ respectively) which we also consider as
the active window.  The coupling of these states to the leads is much
stronger than in the previous case.  Fig.\,9(a) shows the input/output
characteristics for a pulse of length $\tau_p=7.5$ ps.  We see 
that the total currents become equal in the second half of the pulse.
This means that the stationary state is practically reached before
the left contact is switched off.  After that the output current is
delayed with respect to the input current. This delay corresponds to
the propagation of electrons along the sample. We have checked this by
estimating the time needed by one electron having the energy $E_4=2t_L\cos
q$ to travel along the wire; we obtain that the traveling time is around
2 ps which agrees with the numerical results.

The input current has a sharp transient maximum as the left contact
opens, but this behavior is not transferred to the output current which
is rather flat.  Again the relaxation input current has a 'step' after a
fast decay. The occupation numbers of the three states within the bias
window are shown in Fig.\,9(b) and explain the origin of the step.
The two higher states depopulate quite fast in the first half of
the relaxation interval, but at some point their depletion suddenly
slows down.  This suggests that the relaxation rates are governed by
time-dependent tunneling rates depending on the coupling to the leads,
but also on the occupation number of the states. Remark that the bias
window becomes empty before another pulse rises up.

The rise time of the pulse can be varied in experiments (see
Ref. \onlinecite{Naser}).  In our model the rise time depends on the
smoothness parameter $\gamma$ used to create the pulses.  In Fig.\,9(c)
we show the total input and output currents for a pulse with a larger rise
time corresponding to $\gamma=0.25$. The transient maxima in the input
current softens and so does the additional shoulder in the relaxation
current. Note however that during the pulse the output current takes
the same value (around 60 pA) and it is still delayed w.r.t.\,to the input
current.

 Finally we would like to comment on the Coulomb interaction which is neglected here.
 The numerical results show that the occupation number of the levels within the active
 window is quite low for the rather large samples considered (see e.g. Figs.\,6(a) and (b)). 
In this case one expects that the Coulomb interaction between electrons located in 
the active states causes very small changes in the transient currents. On the other hand,
the Coulomb repulsion generated by the inactive states which are fully occupied could 
be important, leading at least to a Hartree shift of the active levels. However, the
relaxation processes and the modulation of the output current should be qualitatively similar.

\section{Conclusions}

We have  analyzed the transient response of a two-dimensional nanosystem
to a sequence of periodic rectangular pulses which modulate the contact
to the source lead.  By solving the generalized non-Markovian master
equation for the reduced density matrix we have been able to discuss the
dependence of the input/output characteristics on the pulse length for
two specific systems: a rather large quantum dot and a narrow quantum
wire. We have considered a pump-and-probe setup in which the contact
to the left lead opens and closes periodically while the right lead is
always connected to the system. When the contact is switched off the
drain current reflects the relaxation processes in the sample. We have
discussed these processes by analyzing the single-particle currents,
the diagonal, and off-diagonal elements of the reduced density matrix.
At a low temperature the phonon effects could be neglected and the main
relaxation processes are back-and-forth tunnelings to and from the leads.

In both cases (the dot and the wire) we have found that the pulse length
can be adjusted such that the shape of the pulse can be reproduced by the
output signal. By increasing the chemical potential of the source lead
the current profile in the output lead develops additional oscillations
related to the relaxation of the higher energy states included in the
bias window.  In the case of a narrow quantum wire a delay of the output
current with respect to the input signal has been obtained.

This study was partly motivated by the recent experiments of Naser
{\it et al.} \cite{Naser} and Lai  {\it et al.} \cite{Lai} Although
those experiments were done with larger and longer pulses our results
qualitatively agree with the reported features of the transient
response. In particular the time-dependent occupation number clearly
show a charging/relaxation behavior as in the work of Lai  {\it et al.}
\cite{Lai} 

\section*{Acknowledgments}
This work was supported by: the Icelandic Science and
Technology Research Program for Postgenomic Biomedicine, Nanoscience
and Nanotechnology; the Computing Center for Design of Materials and Devices,
Icelandic Research Fund grant 090025011; the Research Fund
of the University of Iceland; the Development Fund of the Reykjavik
University grant T09001.   
V.M also acknowledges the hospitality of the Science Institute and the
partial financial support from PNCDI2 programme (grant No. 515/2009) and
grant No. 45N/2009.

\end{document}